# A Review of the Min-Max Approach to the Solution of Relativistic Electron Wave Equation

Sambhu Nath Datta
Department of Chemistry, Indian Institute of Technology – Bombay, Powai,
Mumbai – 400 076, India
Email: sndatta@chem.iitb.ac.in

The variation problem associated with the solution of Dirac's relativistic electron equation is reviewed here. Derivation of the min-max theorem is discussed. A new observation is that the spurious roots of negative energy satisfy a max-min theorem. The min-max principle (MMP) for solution of Dirac equation, extendable to the Dirac-Fock case, is concisely reviewed. MMP for two-electron Dirac-Coulomb equation is discussed. The min-max theorem is physically interpreted for both Dirac and Dirac-Coulomb problems. Applications of MMP are collated in tables. Associated theoretical and computational developments are outlined. Limitations of MMP are spelt out and recent mathematical developments are discussed.

# 1. Introduction

With the advent of Dirac's relativistic electron theory, it was realized that the relativistic wave equation permits negative-energy solutions associated with energy eigenvalues less than or equal to $-mc^2$. This would lead to the hitherto unforeseen circumstance that the normal electron of positive energy would not be stable as it would undergo a spontaneous transition to a negative-energy state, liberating energy of the order of or even greater than $2mc^2$ in the process. This is completely unlike the stability of electrons in the non-relativistic quantum mechanics of Schrödinger. To circumvent this difficulty, Dirac himself proposed that in the neutral state of an atom, the negative-energy states that are infinitely large in number are completely filled. The vacuum state corresponds to an infinite "sea" of electrons in negative-energy states. The absence of one electron from the negative-energy sea would appear as a positively charged hole, and thus Dirac's "hole" theory was born. The charge conjugate wave function $\psi_C = i\gamma_y \psi^*$ satisfies the time-dependent Dirac's equation with the same mass $m$ and the same spin expectation value $<\sigma_D>_C = <\sigma_D>$ but the opposite charge $-e$. This development eventually led to the growth of quantum field theory of relativistic electrons, and then to (relativistic) quantum electrodynamics that is based on the consideration of the quantized fields of electron, positron and photon.

Physicists were unhappy on another issue. Brown and Ravenhall observed that a state of two bound electrons would be degenerate with infinite number of two-electron states with one negative-energy electron (necessarily in a continuum state) and one positive-energy electron (that must belong to the positive energy continuum) [1]. Thus, the two-electron state will, so to say, dissolve into a continuum of degenerate states. The situation is somewhat comparable to, though much more extensive than, the process of auto-ionization on excitation by light. Breit showed that this problem can be avoided by considering only the positive-energy states of one electron, and adding an additional operator (Breit interaction) that must be considered only through the first order in perturbation theory [2]. Breit's work was on the Dirac-Coulomb Hamiltonian, and becomes relevant for a system with more than one electron. Ultimately the idea of the projected interaction was put forward by Sucher, along with the so-called "no-pair" Hamiltonian [3-4]. He also pointed out that the projected Breit interaction can also be used in the SCF process. Nevertheless, the bare one-electron part of the Hamiltonian remains in Sucher's formulation. It would cause no difficulty when the state vector is constrained to remain in the *N*-particle sector of Fock space, but the dilemma of how to identify this sector remains fundamentally unsolved. However, progress has been made by adopting different physical pictures based on the one-electron solutions (Free particle picture and Furry bound state interaction picture) or solutions in the one-electron approximation (Fuzzy picture of Dirac-Fock).

From the viewpoint of quantum mechanical calculation on bound states, Dirac Hamiltonian $\mathbf{H}_D$ presents a lack of variation stability as the energy that can be calculated for the physical ground state is not bounded from below. Early researchers such as Grant [5] and Desclaux [6] were aware of this problem though it hardly affected their work. The reason for the latter was that these authors primarily worked on atomic systems where the orbitals have well-known angular momentum symmetries. Thus, they had to find only the radial functions. In actual calculation, Desclaux followed Hartree's in-out integration method [7] while performing numerical solution of the Dirac-(Hartree)-Fock integro-differential equations, and could obtain radial functions near Dirac-Fock limit. Grant relied on analytical functions, and initially refrained from generating numerical results until a somewhat later time.

There will be no advantage of angular symmetries in a general molecular calculation, and the calculation may undergo a variation collapse [8]. Still, from calculations using carefully chosen basis functions, one rarely finds any collapse. This happens because the calculated energy is much greater than the ground state energy, a basis size truncation effect. Nevertheless, the calculated energy would have error. One option to remove the same error is to estimate the correction to the calculated energy by requiring that the wave function remains orthogonal to the negative-energy functions [8], much like what is done in a non-relativistic valence-electron treatment where one must use pseudopotentials to take care of the core-valence orthogonality. By construction, the pseudopotential was known to have a repulsive contribution [9]. This gave the first hint that a maximization of the calculated energy is required [8].

After the 1980 paper [8] had come out, the variation problem received new attention from different research groups. The use of the squared Dirac operator that should have only positive eigenvalues was suggested [10]. This was an innovative idea, though in 1981 it appeared impractical as the Hamiltonian and its square would generally have different eigenvalues and eigenvectors in a finite basis set. Nevertheless, Kutzelnigg's suggestion eventually pointed out the importance of the right matrix product. Ishikawa and Malli [11] advocated the idea of an effective Hamiltonian for Dirac-Hartree-Fock calculations, which is the right step following ref. 8. Drake and Goldman used discrete $L^2$ basis set to obtain variational results for one-electron atoms [12], while Mark and Schwartz [13] prepared a new representation of the **α.p** operator for use in solving Dirac-type equations by basis set expansion. The choice of basis sets was investigated by Schwarz and Wallmeier who also proposed a quasi-unitary transformation of the Dirac equation in spinor space to obtain a better-behaved matrix representation [14]. Schwarz and Wechsel-Trakowski [15] discussed that the two problems, variation instability and continuum dissolution as discussed in the first two paragraphs of the present article, may affect the accuracy of Dirac-Fock and relativistic configuration interaction (RCI) calculations. Gazdy [16] along with Ladanyi [17] proposed minimizing the least square error in the wave function from a Dirac-type calculation.

A few years earlier, an important development had been made by Rosicky and Mark [18] who had shown that a stationary point can be found from variation of the expectation value $E = \langle \mathbf{H}_D \rangle$ with the operator $\mathbf{\Omega}$ that couples the upper component ($u$) and the lower ($l$) component of the trial spinor by the relation $l = \Omega u$. Solutions of Dirac's equation have (at least) four components. The components $u$ and $l$ are 2-component spinors. One finds the optimized coupling operator $\mathbf{\Omega}_{\pm}^0[u] = (E_{\pm}^0[u] \pm mc^2 - e\Phi)^{-1} c\boldsymbol{\sigma} \cdot \mathbf{p}$, the stationary value $E_{\pm}^0[u]$ being a functional of the trial $u$. Here $\sigma_1$, $\sigma_2$ and $\sigma_3$ are Pauli's spin matrices of rank 2. Furthermore, $E_{\pm}^0[u]$ can be determined by inserting the optimized form of the coupling operator, $\mathbf{\Omega}_{\pm}^0[u]$, in the expression for $\langle \mathbf{H}_D \rangle$. Arbitrary variations of $u$ can lead to the eigenspinors of $\mathbf{H}_D$. Within the framework of the elimination method, the Fock variation of scale in $E_{+}^0[u]$ led to the relativistic virial theorem for the upper component functions [18]. Because of this reason Mark and Schwarz suggested the use of a kinetically balanced basis set [13]. The latter consists of a set of upper component functions $\{u\}$ and a set of lower component functions $\{l\}$ of the form N$\boldsymbol{\sigma} \cdot \mathbf{p}u$, N being a normalizing factor.

It was shown in 1984 that for general atoms and molecules, (1) the extremum $E_{-}^0[u]$ is a minimum ("an upper bound for the energy of negative mass spinors") for a well-behaved (normalizable) $u$ and $E_{+}^0[u]$ is a maximum (with an approximately "orthogonal positive mass spinor") [19]; (2) A "further parametric variation" can lead to a "local minimum for the positive mass spinor," which clearly implied that a constrained-component variation, that is, the variation of $E_+^0[u]$ with $u$ can be used to obtain the discrete eigenvalue spectrum so that a minimum principle exists [19]; (3) It was also demonstrated that the constrained-component variation principle reproduces the Drake-Goldman variation results with the same radial component function for $u$ and $l$ for a hydrogen-like system; (4) Numerical merit of the constrained-component method vis-à-vis the minimal basis LCAS-MS (linear combination of atomic spinors forming a molecular spinor) scheme was assessed. The variation principle was called here as the constrained-component variation principle.

The year 1984 witnessed several other advances. For instance, Stanton and Havriliak [20] argued that when the basis set is kinetically balanced, the calculation is stable with the converged eigenvalue correct through order $v^2/c^2$. However, see Dyall's observation in 1990 on calculations involving a finite Gaussian basis set [21]. In the context of solving Dirac-type equations, Grant and his co-workers considered the problem that the product of two matrices is not equal to the matrix of the operator product in a finite basis set. So, they first rationalized the proper matrix representation of the kinetic energy operator [22] and then completed the calculation. Dyall et al. also derived the criteria for the choice of basis functions for the solution of the Dirac equation from the considerations of matrix products involving the Dirac kinetic energy, and discussed the minimum basis set calculations for the ground state of hydrogen atom [23]. Later, Wood et al. discussed a set of separation theorems associated with the partitioning of the Dirac equation in matrix representation [24], and Grant himself presented a rigorous theorem on the eigenvalue distribution [25]. Since then, the proper matrix representation has become the most popular staple for the computational solution of Dirac-type equations.

In 1986, Talman proposed a min-max principle for solving Dirac's equation in a finite basis expansion method [26], which appears to be more or less equivalent to the "constrained component" variation principle discussed in ref. [19]. The only difference is that the constrained component variation is not limited to a basis expansion but also applicable to a single trial function (analytical and parametric in form). These are strong min-max principles, because both proposals seek the minimum over the upper component of the maximum over the lower component function of the energy functional.

As the bound states have a least eigenvalue, minimization of the stationary value by varying the upper component function $u$ yet preserving the form of the optimized coupling operator leads to the ground state energy $E_g$, and thus we could conclusively show in 1986 and finally in1988 that a min-max theorem exists for solution of Dirac's equation as well as for the Dirac-Fock problem [27]. In the following year, Drake and Goldman who had been working on the variation principle for atoms in the relativistic context discussed the relativistic Sturmian and the finite basis set methods [28]. Four years later in 1992 we could present a form of the min-max theorem for the bound states of the two-electron Dirac-Coulomb equation [29]. Using the positive-energy solution and well-behaved wave functions for the two-electron system, the method becomes free from variational collapse and continuum dissolution. The min-max solution in this case corresponds to the renormalized ground state of quantum electrodynamics, avoiding virtual electron-positron pair(s). Essentially the same min-max theorem was discussed by Kolakowska et al. in 1996 [30]. Whereas coupling operators were considered as parameters of variation in [29], component functions were taken as variation parameters in [30].

Later, Griesmer and Siedentop [31] and Dolbeault et al. [32] added a somewhat elaborate analysis of the strong min-max principle for the one-electron Dirac equation in the language of mathematical spaces. Griesmer and Siedentop considered alternative sets of hypothesis, allowing for operators with Coulomb singularities. Dolbeault et al. discussed the general min-max characterization of the eigenvalues in a gap of the essential spectrum of a self-adjoint unbounded operator.

This article at first reviews the variation problem associated with the solution of Dirac's equation. This is done in Section 2 where the discussion continues on variation stability, Rosicky-Mark variation and its outcomes in the form of a few comments. Next the derivation of the min-max theorem for solving Dirac's equation is given in Section 3. The maximum theorem and the minimum theorem are obtained, algebraic details of minimization are given (in the light of the non-relativistic limit), and finally the min-max theorem is derived. As a new observation, it is shown that the "spurious" roots of negative energy satisfy a max-min principle. The min-max procedure is illustrated and an additional example is given. The min-max principle is concisely reviewed in Section 4 where it is also discussed for the two-electron Dirac-Coulomb equation. A physical interpretation of the min-max theorem is given. Besides, different physical and mathematical applications of the min-max principle are mentioned. These examples are gathered together in tabulated forms. Associated theoretical and computational developments are briefly mentioned. In the final section, namely, Section 5, limitations of the min-max procedure are discussed. This section is closed with a brief discussion on the aspects of the variation theorem based methods of solution of Klein-Gordon equation that gives the kinematics for zero-spin boson.

## 2. The Variation Problem

### 2.1. Variation Stability [8]

Consider Dirac's Hamiltonian for a spin-1/2 particle of charge $e$ and mass $m$, and moving under the influence of an (external) scalar potential $e\Phi$,

$$\mathbf{H}_D = c\boldsymbol{\alpha} \cdot \mathbf{p} + \beta mc^2 + e\Phi \tag{1}$$

where $\alpha$ and $\beta$ are standard matrices of rank 4. For a meaningful attractive potential, $-mc^2 \ll e\Phi < 0$ at every point in space except in the neighbourhood of a finite number of singularities such that $-mc^2 < \langle e\Phi \rangle < 0$ holds for all normalizable trial functions. The corresponding eigenvalue spectrum is divided into three parts: (i) the positive-energy continuum with $\infty > E_+ > mc^2$, the corresponding set of eigenspinors $\{\psi_+\}$ representing the scattered states of the particle; (ii) discrete energy spectrum with $mc^2 \geq E > 0$, the corresponding eigenspinors being the typical bound state wave functions; the physical ground state corresponds to the least discrete eigenvalue $E_g$ and the corresponding eigenfunction $\psi_g$; (iii) the negative energy continuum with $-mc^2 > E_- > -\infty$ corresponding to spinors $\{\psi_-\}$ which, by charge conjugation operation, convert into the positive-energy continuum states of the antiparticle.

The eigenfunctions are (at least) 4-component spinors. A trial spinor can be written as

$$\psi = \begin{pmatrix} u \\ l \end{pmatrix}, \tag{2}$$

where the upper and lower components *u* and *l* are 2-component spinors of the Pauli type. Each component (*u* and *l*) represents a net (spin plus orbital) angular momentum state, and together they represent the total angular momentum as well as the charge degree of freedom. The explicit form of the charge conjugate wave function is

$$\psi_C = i\gamma_y \psi^* = \begin{pmatrix} i\sigma_y l^* \\ -i\sigma_y u^* \end{pmatrix}. \tag{3}$$

Consider the Dirac Hamiltonian matrix written in the basis of known 4-component spinors of positive energy. Each basis spinor is written in terms of well-behaved upper and lower component functions. A diagonalization would result in a discrete spectrum. The minimum eigenvalue is $\varepsilon_0$ corresponding to the eigenvector $\psi_0$. The latter is not necessarily orthogonal to the normalized $\psi_-$'s. Mixing of the negative-energy wave functions would generally contribute to a lower than expected energy value, and may even make $\varepsilon_0 < E_g$ (variation collapse). Let us write the optimized spinor $\psi_0$ as

$$\psi_0 = \psi_g + \mathbf{P}_- \psi_0, \tag{4a}$$

$$\mathbf{P}_- = \mathop{\mathbf{S}}_{\psi_-} |\psi_-\rangle\langle\psi_-| \tag{4b}$$

being the negative-energy projector. It is easy to show

$$(\mathbf{H}_D + \mathbf{V}^{NEPP})\psi_0 = E_g \psi_0 \tag{5}$$

where $\mathbf{V}^{NEPP}$ is the negative-energy pseudopotential,

$$\mathbf{V}^{NEPP} = -\mathbf{P}_- \mathbf{H}_D - \mathbf{H}_D \mathbf{P}_- + \mathbf{P}_- \mathbf{H}_D \mathbf{P}_- + E_g \mathbf{P}_-. \tag{6}$$

Because $\psi_-$ is an eigenspinor of $\mathbf{H}_D$, the operator $\mathbf{P}_-$ commutes with the same Hamiltonian and one finds $\mathbf{V}^{NEPP}$ in the Phillip-Kleinman form

$$\mathbf{V}^{NEPP} = \mathop{\mathbf{S}}_{\psi_-} |\psi_-\rangle (E_g - E_-) \langle\psi_-|, \tag{7}$$

with a manifestly repulsive contribution of the order of $2mc^2 \langle \mathbf{P}_- \rangle$ to energy. The operator $(\mathbf{H}_D + \mathbf{V}^{NEPP})$ is bounded from below, that is, $\langle \mathbf{H}_D + \mathbf{V}^{NEPP} \rangle \geq E_g$ [8], in agreement with the general assessment for pseudopotentials by Weeks et al. in ref. [9].

**2.2. Rosicky-Mark Variation** [18]

One may write $l = \Omega u$ and $\varepsilon = \langle \mathbf{H}_D \rangle_\psi$ to find

$$\begin{aligned}(u,[\varepsilon[u] - mc^2 - e\Phi]u) \\ +(\Omega u,[\varepsilon[u] + mc^2 - e\Phi]\Omega u) - c[(\Omega u, \boldsymbol{\sigma}\cdot\mathbf{p}u) + (u, \boldsymbol{\sigma}\cdot\mathbf{p}\Omega u)] = 0\end{aligned} \tag{8}$$

such that

$$\begin{aligned}\varepsilon[u] = [(u,u) + (\Omega u, \Omega u)]^{-1} \Big[ mc^2[(u,u) - (\Omega u, \Omega u)] \\ + [(u, e\Phi u) + (\Omega u, e\Phi \Omega u)] + c[(\Omega u, \boldsymbol{\sigma}\cdot\mathbf{p}u) + (u, \boldsymbol{\sigma}\cdot\mathbf{p}\Omega u)] \Big].\end{aligned} \tag{9}$$

In all these expressions, the bracket (…) represents a scalar product. A variation of $\Omega u$ as $\Omega u \rightarrow \Omega u + \delta(\Omega u)$ in (9) is accompanied by a change of the energy expectation value as $\varepsilon \rightarrow \varepsilon + \delta\varepsilon + \delta^2\varepsilon$, from which we get the first order energy change

$$\begin{aligned}\delta\varepsilon = -[(u,u) + (\Omega u, \Omega u)]^{-1} \Big[ (\delta\Omega u,[\varepsilon + mc^2 - e\Phi]\Omega u) + c(\delta\Omega u, \boldsymbol{\sigma}\cdot\mathbf{p}u) \\ + (\Omega u,[\varepsilon + mc^2 - e\Phi]\delta\Omega u) + c(u, \boldsymbol{\sigma}\cdot\mathbf{p}\delta\Omega u) \Big]\end{aligned} \tag{10}$$

that is to be used now, and the second order infinitesimal

$$\delta^2\varepsilon = -\left[(u,u)+(\Omega u,\Omega u)\right]^{-1}\left[\left[(\delta\Omega u,u)+(u,\delta\Omega u)\right]\delta\varepsilon+(\delta\Omega u,[\varepsilon-e\Phi]\delta\Omega u)\right] \quad (11)$$

for a later usage.

When $\delta\varepsilon[u] = 0$ for arbitrary $\delta(\Omega u)$, one finds the optimum form of the coupling operator

$$\Omega^0(u) = (\varepsilon^0[u] + mc^2 - e\Phi)^{-1}c\boldsymbol{\sigma}\cdot\mathbf{p} \quad (12)$$

with the optimized lower component function for the specific upper component function $u$ as

$$l = (\varepsilon^0[u] + mc^2 - e\Phi)^{-1}c\boldsymbol{\sigma}\cdot\mathbf{p}u \quad (13)$$

at points where the inverse operator exists. At points where it does not exist the upper component vanishes to zero value. Equation (13) is in principle identical with the second line of Dirac's equation with the expectation value substituting for the eigenvalue,

$$c\boldsymbol{\sigma}\cdot\mathbf{p}u - mc^2 l + e\Phi l = \varepsilon^0[u]\, l. \quad (14)$$

When the operator $\Omega^0(u)$ is inserted into Eq. (9) one finds that the stationary value $\varepsilon^0[u]$ satisfies the Rosicky-Mark equation

$$(u,[\varepsilon^0[u]-mc^2-e\Phi]u) = c^2(u,\boldsymbol{\sigma}\cdot\mathbf{p}[\varepsilon^0[u]+mc^2-e\Phi]^{-1}\boldsymbol{\sigma}\cdot\mathbf{p}u). \quad (15)$$

By making $\varepsilon^0[u]$ stationary for a scale variation of the component $u$, Rosicky and Mark derived the relativistic virial theorem for the bound state in terms of only the upper component of the Dirac wave function. They also derived general formulae for the relativistic electron moving in an electrostatic central field. Furthermore, using series expansions, a nonrelativistic approximation to the virial theorem up to second order in the fine-structure constant was obtained by these Authors.

## 2.3. Comments

*Comment 1:* Alternatively, by the use of Fock's variational argument (the principle of least action), making $\varepsilon^0$ stationary for *all* arbitrary variation $\delta u$ leads to the complete equation of Dirac,

$$mc^2 u_i + e\Phi u_i + c\boldsymbol{\sigma}\cdot\mathbf{p}l_i = \varepsilon_i u_i \quad (16a)$$
$$c\boldsymbol{\sigma}\cdot\mathbf{p}u_i - mc^2 l_i + e\Phi l_i = \varepsilon_i l_i \quad (16b)$$

where $u_i$ and $l_i$ are the components of the eigenspinor $\psi_i$ corresponding to the discrete eigenvalue $\varepsilon_i$.

*Comment 2:* The derivation by Rosicky and Mark rests upon the assumption of a normalizable spinor, and so it obviously leads to a discrete energy spectrum and the corresponding bound state eigenspinors. However, when the so-called box normalization is considered (as it is routinely practised in different theories of solid state physics, and for quantizing the electromagnetic field), the derivation of equations (16) can be extended to the realm of positive-energy as well as the negative-energy continuum solutions in the limit of infinite volume.

*Comment 3:* Finally, for a well-behaved (normalizable) $u$, the Rosicky-Mark equation (15) can be written as

$$\varepsilon^0[u]^2 = m^2c^4 + c^2<p^2>_u + (\varepsilon^0+mc^2)<e\Phi>_u + \frac{c^2<\boldsymbol{\sigma}\cdot\mathbf{p}e\Phi\boldsymbol{\sigma}\cdot\mathbf{p}>_u}{(\varepsilon^0+mc^2)} + \frac{c^2<\boldsymbol{\sigma}\cdot\mathbf{p}e^2\Phi^2\boldsymbol{\sigma}\cdot\mathbf{p}>_u}{(\varepsilon^0+mc^2)^2} + \ldots$$

$$(17)$$

that has at least the pair of solutions [19]

$$\varepsilon_+^0[u] \simeq +mc^2 + \left( \frac{<p^2>_u}{2m} + <e\Phi>_u \right) + \frac{<\sigma\cdot pe\Phi\sigma\cdot p>_u}{4m^2c^2} + \frac{<\sigma\cdot pe^2\Phi^2\sigma\cdot p>_u}{8m^3c^4} + ...$$

and                                                                                                     (18)

$$\varepsilon_-^0[u] \simeq -mc^2 - \frac{<p^2>_u}{2m} + ...$$

For $<e\Phi>_u \approx -\frac{<p^2>_u}{m} \approx -\alpha^2 Z^2 mc^2$ where $\alpha$ is fine structure constant ($\alpha = e^2/\hbar c$), a relation reminiscent of the non-relativistic virial theorem and the expression for atomic energies, one obtains $\varepsilon_+^0[u]$ in the range $+mc^2 > \varepsilon_+^0[u] > 0$ for $\alpha Z < 1$. It is also evident that $-mc^2 > \varepsilon_-^0[u]$. We shall find these inequalities very useful in the following discussion.

***Comment 4:*** As $e\Phi \to 0$, the free particle limit $\varepsilon^0[u]^2 \to m^2c^4 + c^2 <p^2>_u$ is retrieved from (17).

## 3. Derivation of Min-max Theorem

When $\varepsilon$ is stationary for variation of $l = \Omega u$, one finds from (11) the curvature

$$\delta^2\varepsilon = -\left[(u,u)+(\Omega u,\Omega u)\right]^{-1}(\delta\Omega u,[\varepsilon^0 - e\Phi]\delta\Omega u). \tag{19}$$

### 3.1. The Maximum

Suppose that the upper component function has a specific analytical form, or that it has been expanded in a finite basis set of two-component normalizable spinors. It is also quite realistic in the sense that it closely resembles an approximate wave function of the non-relativistic treatment. Furthermore, the lower components are prepared such that the coupling operator approximates $(2mc^2 - e\Phi)^{-1}c\sigma\cdot\mathbf{p}$. Then one rarely finds an energy expectation value $\varepsilon$ that is less than $\varepsilon_g$ [19]. For a meaningful, attractive Coulomb potential that is normally encountered in atomic and molecular physics, one finds using a realistic four-component trial spinor the result $mc^2 \sim \varepsilon \gg 0$ for $\alpha Z \ll 1$. Therefore, $\delta^2\varepsilon$ in (19) is certainly negative at points corresponding to the positive-energy solutions of the Rosicky-Mark equation (15) and its expanded form (17). The stationary value $\varepsilon_+^0[u]$ from the constrained-component variation is a maximum [19, 27],

$$\max_{\Omega u} <\mathbf{H}_D>_{u,\Omega u} = \varepsilon_+^0[u]. \tag{20}$$

### 3.2. The Minimum

As long as the relation $\varepsilon^0[u] \gg -mc^2$ is satisfied for the normalizable $u$, arbitrary variation of $\varepsilon^0[u]$ with $u$ will lead to one of the discrete eigenvalues. This indicates that the four-component spinor can be expanded in the basis of bound state eigenspinors. This is also corroborated by the first expansion in (18). Therefore, one obtains a minimum theorem [27]

$$\min_u \varepsilon_+^0[u] \geq E_g . \tag{21}$$

While discussing the constrained component variation procedure, the maximization step was described in the context of a Ritz variation in the basis of the separate upper and lower component spinors, whereas the minimization process was illustrated using the Drake-Goldman spinor. See ref. [19] for details.

### 3.3. Algebraic details

The minimum theorem (21) can also be understood in algebraic details by starting from (15) and considering the changes in the positive-energy solutions as $u^0 \to u^0 + \delta u^0$ and $u^{0\dagger} \to u^{0\dagger} + \delta u^{0\dagger}$ such that $\varepsilon^0 \to \varepsilon^0 + \delta\varepsilon^0 + \delta^2\varepsilon^0$. At the stationary point $\delta\varepsilon^0 = 0$ and one finds

$$[(u^0,u^0)+(l^0,l^0)]\,\delta^2\varepsilon^0 = \\ -(\delta u^0, \left[\varepsilon^0[u^0] - mc^2 - e\Phi - c^2\sigma\bullet p\ [\varepsilon^0[u^0] + mc^2 - e\Phi]^{-1}\sigma\bullet p\right]\delta u^0). \quad (22)$$

To the lowest order in $v/c$, one can simplify the first term in $\delta^2\varepsilon$ as

$$[(u^0,u^0)+(l^0,l^0)]^{-1}\varepsilon^0[u^0] = (u^0,\left[1+\frac{p^2}{4m^2c^2}\right]u^0)^{-1}\varepsilon^0[u^0] = \varepsilon_{nonrel}[\varphi^0], \quad (23)$$

where $\varphi^0 = \left(1-\frac{p^2}{8m^2c^2}\right)u^0$ is the wave function in the non-relativistic limit and $\varepsilon_{nonrel}$ is the corresponding energy that is correct through order $v^2/c^2$.

The deviation $\delta\varphi^0$ that corresponds $\delta u^0$ can be written as

$$\delta\varphi^0 = (u^0,\left[1+\frac{p^2}{4m^2c^2}\right]u^0)^{-1/2}\delta u^0. \quad (24)$$

The rest of the terms in the right side of (22) can be simplified as

$$(u^0,\left[1+\frac{p^2}{4m^2c^2}\right]u^0)^{-1}(\delta u^0,\left[mc^2 + e\Phi + c^2\sigma\bullet p\ [\varepsilon^0[u^0] + mc^2 - e\Phi]^{-1}\sigma\bullet p\right]\delta u^0)$$

$$= (u^0,\left[1+\frac{p^2}{4m^2c^2}\right]u^0)^{-1}\frac{(\delta u^0,\left\{mc^2 + e\Phi + \frac{p^2}{2m}+...\right\}\delta u^0)}{(\delta u^0,\delta u^0)}(\delta u^0,\delta u^0) \quad (25)$$

$$= (u^0,\left[1+\frac{p^2}{4m^2c^2}\right]u^0)^{-1}\varepsilon^0[\delta u^0](\delta u^0,\delta u^0)$$

$$= \varepsilon_{nonrel}[\delta\varphi^0](\delta u^0,\delta u^0).$$

Therefore, from (22),

$$\delta^2\varepsilon^0 = -(\varepsilon_{nonrel}[\varphi^0] - \varepsilon_{nonrel}[\delta\varphi^0])(\delta u^0,\delta u^0). \quad (26)$$

However, $\varepsilon^0_{nonrel}[\delta\varphi^0] > \varepsilon^0_{nonrel}[\varphi^0_0]$, $\varphi^0_0$ being the non-relativistic ground state. This is well known at the Schrödinger level, and the proof can be easily extended to the non-relativistic limit from (18) on the ground that $v^2/c^2 \ll 1$. Thus for the positive-energy solution of Eq. (17), the second derivative is positive ($\delta^2\varepsilon^0 > 0$) at the stationary point that leads to the minimum theorem (21).

The optimal trial function $u^0_0$ is related to the non-relativistic ground state wave function $\varphi^0_0$ by $\varphi^0_0 = \left(1-\frac{p^2}{8m^2c^2}\right)u^0_0$.

### 3.4. The Min-Max Theorem

Equations (20) and (21) together represent the min-max theorem [27]

$$\min_{u} \left( \max_{\Omega u} <\mathbf{H}_D>_{u,\Omega u} \right) \geq E_g. \tag{27}$$

It is a rigorous theorem under the restrictions that have been stipulated, and it gives a strong min-max principle. Kutzelnigg [33], however, criticized it, apparently thinking that the variation of $u$ and $l$ had been independently made, and went on stressing the point of constraining the components. He was also apparently not aware of the earlier developments in reference [8] about uplifting (maximizing) energy and in reference [19] about constrained component variation, and particularly overlooked the importance of the derivation of Dirac's equation from what we know as Fock's principle of least action [18]. Rosicky and Mark had been apparently satisfied with the derivation of the relativistic virial relation from a constrained component variation of only the scale factor [18], and had not gone forward towards a general variation principle. Still, as mentioned earlier, they were the earliest investigators in this area. Later, Mark went on to prepare a modified matrix representation of the odd term in Dirac's operator [13] and then worked out very accurate, kinetically balanced, GTO basis sets for Dirac-Fock calculations on first- and second-row atoms [34]. Stable near Dirac-Fock calculations on the same elements were also done using Gaussian basis sets [35-36] that could be prepared from the non-relativistic Partridge basis [37] and using coupling operators of the form $\Lambda\sigma\cdot p$, the parameter $\Lambda$ ($\approx 1/2c$) being optimized by variation.

### 3.5. Max-min theorem for the spurious root

Comparing (19) one finds that the spurious root $\varepsilon_-^0[u]$ ($< -mc^2$) is a minimum as long as $<e\Phi>_{\delta\Omega u} > -mc^2$:

$$\min_{\Omega u} <\mathbf{H}_D>_{u,\Omega u} = \varepsilon_-^0[u]. \tag{28}$$

However, $\varepsilon_-^0[u]$ is bounded from above and not from below. Hence a variation with respect to $u$ can lead towards a maximum value:

$$\max_{u} \varepsilon_-^0[u] \to -mc^2 + 0-. \tag{29}$$

The maximum $\varepsilon_{-,\max}^0 = -mc^2$, however, cannot be achieved from a non-trivial negative-energy solution. It can be realized only for the trivial spinor that has the upper component as the null column vector and the lower component as the column vector of norm unity, and when there is no scalar potential ($e\Phi$ is zero at every point in space). In any case one gets the max-min theorem as the inequality

$$\max_{u} \min_{\Omega u} <\mathbf{H}_D>_{u,\Omega u} < -mc^2. \tag{30}$$

### 3.6. Illustration

Variation collapse on both sides of the maximum was demonstrated as early as in 1988 for Dirac equation on $H_2^+$ and Dirac-Fock equations on $H_2$ when the coupling operators in the basis spinors of the LCAS-MS scheme are parametrically written in the form of $\Lambda\sigma\cdot p$ [27]. A minimum of the maximum values produced using the positive root of energy gives rise to a "shower" of energy trajectories on the two sides of the minimum along the coupling operator degree of freedom. This is illustrated in Figure 1. The minimal basis $1s\sigma_{1/2}$ function prepared from the 1s STOs with exponent $\zeta$ was used to write the upper component $u$. The lower component was prepared by applying the coupling operator $\mathbf{\Omega} = \zeta^{-1}[(1-\gamma)/(1+\gamma)]^{-1/2}\boldsymbol{\sigma} \cdot \mathbf{p}$ on $u$, where $\gamma = (1 - \alpha^2\kappa^{-2}Z^2)^{-1/2}$. The factor of inverse exponent keeps the exponent dependence of the lower component at par with that of the upper component. The reason for using this basis was that it had been used to generate the correct energy eigenvalues and eigenspinors at the united atom limit and the separated atom limit [19]. The coupling operator was varied by changing $\kappa$ which gave rise to an energy maximum for each STO exponent. The surrounding collapses showed an excess stability for the $1\sigma_{1/2}$ spinor [27].

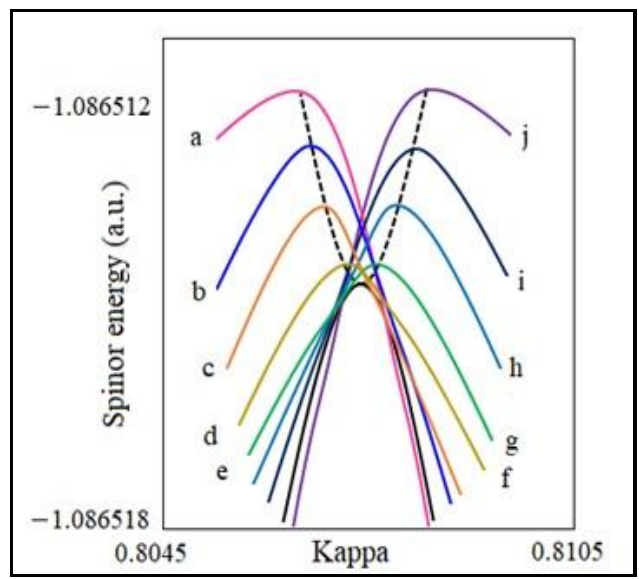

**Figure 1.** Shower diagram illustrating the min-max principle for the minimal basis calculation on $1\sigma_{1/2}$ spinor of $H_2^+$ molecule-ion with internuclear separation of 2.0 Bohr. Variation of the $u$-$l$ coupling operator produces a maximum for each exponent $\zeta$ of the STOs used to form the upper component $u$. The upper component is varied as $\zeta$ decreases from (a) 1.2410 to (j) 1.2365 in alphabetic order in steps of 0.0005, producing a minimum of the maximum values. (Reproduced and modified from ref. [27]).

### 3.7. Additional Example

For He atom, the basis functions were chosen by updating the non-relativistic basis set (1s, 2s, 3s and 4s STOs with $\zeta=2.0$) to a set of 4-component spinors with a uniform $\Lambda$. The uniform value was found to be $0.4999734c^{-1}$. The calculated wave functions – both non-relativistic and relativistic – progressively improved with the increase in basis size, and produced better 1s orbital energy. The difference ($\varepsilon_{rel}-\varepsilon_{nonrel}$) was found to be $-3.1\times10^{-5}$ a.u., slightly greater than the amount $-3.4\times10^{-5}$ a.u. predicted by numerical techniques. In spite of the rather small basis set, the total energy difference ($E_{rel} - E_{nonrel}$) converged to a value $-13.8\times10^{-5}$ a.u. for $\zeta=2.0$, and $-10.1\times10^{-5}$ a.u. for $\zeta=1.6875$. These values are comparable to the numerical Dirac-Fock energy difference $-12.5\times10^{-5}$ a.u., a little less or greater depending on how well the relativistic wave function is described near the nucleus vis-à-vis away from the nucleus.

These examples serve to illustrate only the min-max theorem. For quantitative results, one ought to consider explicit calculations in references 25, 34, 36, and elsewhere.

**4. Min-Max Principle (MMP) and Applications**

**4.1. MMP**

The min-max principle directly stems from the minimization of the solution $\varepsilon_+^0[u]$ of the Rosicky-Mark equation [18] as discussed in references [19] and [27]. It was also obtained in [26] while considering the physical roots of Drake-Goldman equation [12] for discrete functions. In the first stage, energy is made stationary for variation of the lower component of the trial spinor, while the functional form of the upper component remains unchanged. (The upper component actually changes by a constant factor as the normalization constant alters with variation of the lower component). The second stage involves variation of the stationary points (maxima) that are associated with different upper component functions. This technique was initially suggested in ref. [19], though Talman coined the name MMP and provided numerical evidence [26] while overlooking ref. [19]. It must be understood that whereas in references [18], [19] and [27] one varies the coupling operator $\Omega$, Talman always varies the lower component $l$ [26], the effect being the same in case the component $l$ is expressible as the result of application of operator $\Omega$ on the upper component $u$, while the overall 4-component spinor is of norm unity.

**4.2. MMP in Dirac-Coulomb treatment**

A similar repetition of the min-max treatment has been observed for the two-electron Dirac-Coulomb equation. In 1992, variation of the expectation value of $\mathbf{H}_{DC}$ with respect to three different coupling operators $\Omega_+$, $\Omega_-$ and $\omega$, and one primary two-electron super-component function either $\Phi_1(\mathbf{r}_1,\mathbf{r}_2)$ or $\Phi_3(\mathbf{r}_1,\mathbf{r}_2)$, led to the theorem [29]

$$\min_{\substack{\Phi_1 \\ (\text{or } \Phi_3)}} \max_{\omega} \max_{\Omega_-} \max_{\Omega_+} <\mathbf{H}_{DC}>_\Psi \geq E_0 \qquad (31)$$

where $\Phi_2 = \Omega_+\Phi_1$, $\Phi_4 = \Omega_-\Phi_3$, $\Phi_3 = \omega\Phi_1$, and the $\Phi_i$'s ($i = 1, 2, 3, 4$) are the four super-components of $\Psi$: $\Psi^\dagger(\mathbf{r}_1,\mathbf{r}_2) = [\Phi_1^\dagger \ \Phi_2^\dagger \ \Phi_3^\dagger \ \Phi_4^\dagger]$. Here, $E_0$ is the energy of the physical ground state. One may also identify it as the least value among the discrete eigenvalue spectrum of Sucher's no-pair Hamiltonian.

In ref. [30] one finds

$$\min_{\Phi_1} \left( \max_{\Phi_3} \max_{\Phi_2} \max_{\Phi_4} <\mathbf{H}_{DC}>_\Psi \right) = \varepsilon \qquad (32)$$

where $\varepsilon$ is one of the matrix eigenvalues for positive energy. No reference was given to [29]. This prompted us to discuss the general variable operator technique leading to the min-max theorem to remove any existing confusion between variation of a coupling operator and a subsidiary function that can be obtained from the coupling [38].

Mathematicians have relatively recently done investigations on operators, their self-adjoint nature and other properties, which is briefly discussed in the next section.

### 4.3. Physical Interpretation

In general, the maximization technique decreases the weight of the negative-energy eigenspinors in the trial positive-energy solution to ultimately obtain the optimized solution with energy $\varepsilon_+^0$. A transition of a particle from a positive-energy state to a negative-energy one is accompanied by the liberation of energy of the order of $2mc^2$. In Dirac's hole theory, however, this process must be blocked as the negative energy sea is completely filled. The process of blocking the transition increases the energy of the electron. So Dirac's hole theory already stipulates the maximum step to be carried out.

In QED picture, the said transition is equivalent to the destruction of a pair of particle and antiparticle, both in positive-energy states. The destruction process evolves energy of the order of $2mc^2$, whereas the opposite process of creation of a pair increases the energy of the system. The creation amounts to transferring an electron from a negative-energy state to a positive-energy orbital. The trial spinor generally contains some contribution from the negative-energy eigenspinors which implies that the trial state vector has a (negative) contribution from particle-antiparticle pairs. The creation process nullifies the negative contribution. The latter is progressively reduced (in amplitude) during the step of energy maximization until it stops as the trial vector reaches the 1-particle sector (becomes devoid of the pair terms). Thus the maximization step described by Eq. (20) accounts for a progressive removal of unphysical contributions from the trial vector. However, in QED it is normal to avoid the pair terms in the one-electron Hamiltonian operator, that is, to work with the diagonal operator in normal ordered form. If one knows the diagonal operator, then one has already solved the eigenvalue problem.

For the two-electron Dirac-Coulomb equation, it was shown in ref. [29] that the min-max principle leads to a result safe from both variation collapse and continuum dissolution. Furthermore, an analysis showed that the second order level shift arising from the Pauli blocking of virtual pair is inherent in the procedure, and the min-max solution represents the renormalized ground state of QED. See Figure 2 for the working of min-max technique, and Figure 3 for the Pauli blocking of diagrams in QED.

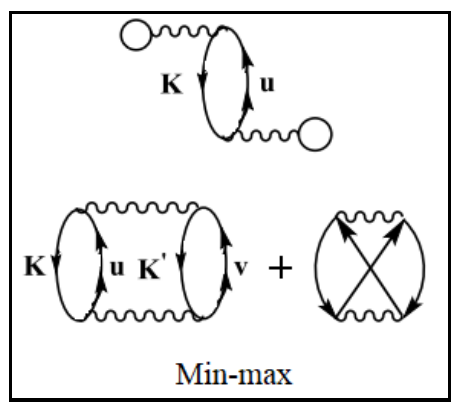

**Figure 2.** Second order diagrams representing the admixing of deexcited configurations during the min-max maximization procedure. A negative-energy electron line (u or v) proceeds upward, and a hole line (**K** or **K′**) runs downward. The second order correction to energy is positive in this case as the denominator is positive. (Reproduced and modified from ref. [29]).

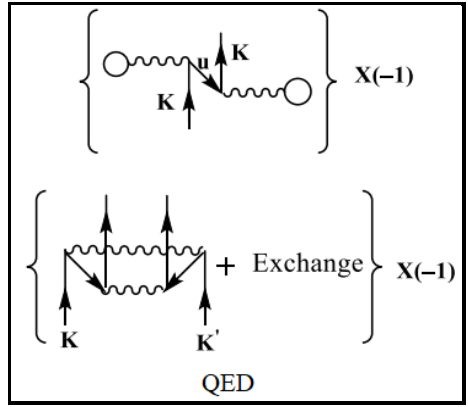

**Figure 3.** Second order diagrams representing the contribution of virtual pairs in QED. A positron line (u or v) runs downward, and an electron line (**K** or **K′**) proceeds upward. The second order correction to energy is negative as usual, but these diagrams are blocked by Pauli's exclusion principle. The net contribution is positive. (Reproduced and modified from ref. [29]).

Numerical comparisons were made with the relativistic configuration interaction (RCI) methods for He-like systems. The Pauli blocking energy has been calculated as the energy difference from the two forms of RCI, using all-energy spinors (AE-RCI) and positive-energy spinors (PE-RCI). It was also formulated using QED in leading order, (equation 98 in reference 29), and estimated for comparison with the calculated ΔRCI values for various basis size and various $Z$. The computed ΔRCI and the QED estimates were quite comparable to each other for $\alpha Z \leq 0.1$, and as expected, deviations increasingly occurred at higher $Z$ values (Table 3 in reference 29). To conclude, the min-max technique makes a good case for eliminating the effect of virtual electron-positron pairs.

Kolakowska et al. have presented variational calculations on the ground state and $n = 2$ complex of states of He-like systems using small number of basis functions and varying non-linear parameters [30]. Of course, they have discussed the advantages of gaining stability and avoiding continuum dissolution. However, they have apparently not detected any QED correction.

**4.4. Applications**

So far application of the MMP has been studied only for a few types of extremely simple systems. Alternative techniques based on the matrix representation procedure (due to Grant's group and the group of Mark and Schwarz) have turned out to be more versatile for quantum chemical calculations, though these are generally accurate in the limit of an infinitely large, (rather, a complete), basis set. Applications of MMP and alternative techniques include (i) investigations of Dirac equation on one-electron atom or atomic ions [12-14, 19, 23, 26, 28, 31-33, 39-41] and one-electron molecule-ions [19, 27, 42], (ii) sample Dirac-Fock calculations on two-electron atoms [27, 29-30] and molecules [27, 35], (iii) general Dirac-Fock and many-body calculations on many-electron atoms [22-25, 29, 34-36, 44] and molecules [8, 11, 13-14, 27, 35, 43], (iv) Dirac-Coulomb calculations on two-electron atoms [29-30], and other related topics including negative-energy corrections [8, 11, 43], virial relations [18, 39], relativistic extension of Hohenberg-Kohn theorem while using the projected Hamiltonian of Sucher and the ensuing density functional treatment [40].

For instance, it is not easy to prescribe a DFT even with the projected Hamiltonian. Consider the hydrogen atom: let the upper component of the trial spinor be defined by $s_n$ and the lower component in terms of $p_{nx}$, $p_{ny}$ and $p_{nz}$ with the common multiplicative factor $\lambda_n$. All four functions are characterized by the same radial function typical of the STO with $l=0$ and principal quantum number $n$. The expectation value $<\mathbf{H}_D>$ can be maximized by varying $\lambda_n$. A follow-up minimization of the maximum value by changing the exponent gives for each $n$ the same energy $E_{1s}= mc^2(1-\alpha^2 Z^2)^{1/2}$ that is the least value among the discrete eigenvalues. Figure 4 shows the corresponding (optimized) radial densities. This apparently negates the Hohenberg-Kohn approach. However, there is a catch – the coupling operator (the lower component) is more or less optimal only for $n = 1$, and in other cases it is written down in an a very approximate way. If the MMP is followed with utmost care for the maximum step, such a fallacy would not occur. For a correct relativistic DFT one must have a density that corresponds to the maximum values for variation of the involved lower components.

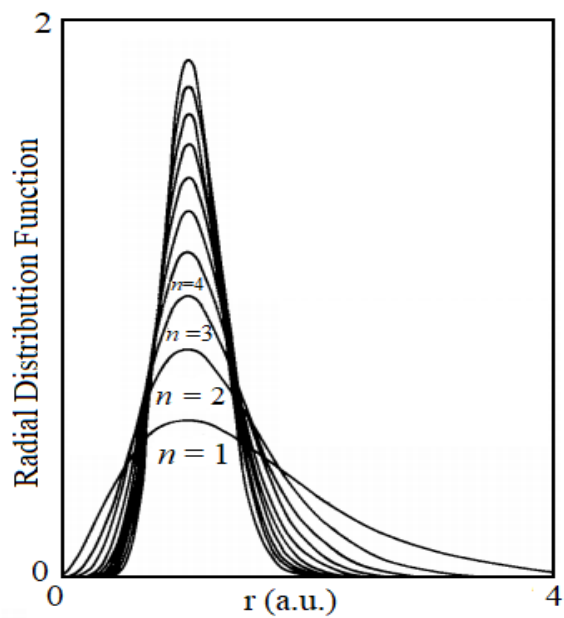

**Figure 4.** Radial densities that yield the same expectation value that is equal to the least discrete eigenvalue $E_{1s}$ of Dirac Hamiltonian for the hydrogenic electron. (Reproduced and modified from ref. [40]).

Table 1, Table 2 and Table 3 list the exemplary systems and phenomena studied and the corresponding references. In particular, Table 1 collates early references for the collapse-free solution of Dirac equation for hydrogen-like systems and one-electron molecule-ions, and Table 2 does the same for Dirac-Fock equations on atoms and molecules. Table 3 cites on the general Dirac-Fock and many-body calculations and Dirac-Coulomb calculations on two-electron atoms.

A due recognition must be given to the contributions made by earlier researchers of relativistic molecular effects. Pavlik and Blinder [45] examined the ground state of the hydrogen molecular ion in 1967 by straight-forwardly applying the variation theorem. They considered trial functions of symmetry similar to that of the nonrelativistic molecular orbitals. Relativistic correction to the electronic energy was estimated to be $-7\times10^{-6}$ a.u. In 1969, Luke et al. [46] reduced Dirac equation to a perturbed Schrödinger equation and determined the first order correction to the $1s\sigma_g$ energy (about $-1\times10^{-5}$ a.u.). A pioneering work was done by Malli and Oreg, who established the linear combination of atomic spinors as molecular spinors (LCAS-MS) scheme to formulate Dirac-Fock calculations on close-shell and open-shell molecules [47-51]. This methodology is often followed today for a general molecular calculation along with the techniques to provide safety against variation collapse. Pyykkö and Desclaux proposed the numerical Dirac-Fock one-centre calculations [52]. This approach becomes computationally tractable only for diatomic molecules and small, highly symmetric polyatomic molecules. Finally, Lee and McLean also reported a general molecular calculation in 1982 [53]. These approaches directly did not aid in formulating the MMP, instead they opened up our eyes for the general molecular methodologies. These contributions have been collated in Table 4.

Miscellaneous applications include the relativistic virial theorem, negative-mass corrections, and relativistic extension of Hohenberg-Kohn theorem. Reference 41 offers several interesting examples like the construction of spinors of s symmetry from STO bases, how to find the continuum state solutions, and derivation of the lowest eigenvalue of the discrete spectrum for each angular momentum. The dilemma in Figure 4 is clarified in Figure 5 where spinors of s symmetry prepared from STO bases have been explicitly treated by MMP. There is no misunderstanding here as regards the energy corresponding to different principal quantum numbers. As in ref. [27], here too the variation collapse has been clearly demonstrated.

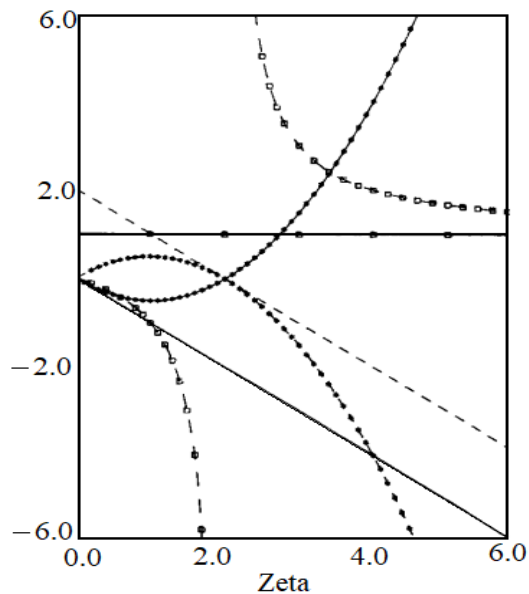

**Figure 5.** Variation of $(\varepsilon_\pm^0 \mp mc^2)$ and $<e\Phi>_\pm$ in atomic units, and $n_\pm$ with the exponent $\zeta$ involved in the STO-based trial spinor representing the electronic state in hydrogen atom: $(\varepsilon_\pm^0 \mp mc^2)$ in atomic units (*****); $<e\Phi>_\pm$ in atomic units (——— and - - -); and $n_+ =1$ (horizontal line with squares), $n_-$ (curved lines with squares and singularity at $\zeta=2.0$). (Reproduced and modified from ref. [41]).

To the knowledge of the present author, two scientific reviews on the variation analysis have been reported, one in 1986 [25] and another in 1994 [54]. General operator properties were investigated in ref. [38], where it was considered that the operator may not be self-adjoint. Also, it was explicitly shown that by varying the *u-l* coupling operator one removes all negative-energy spinor contributions to the trial spinor. These references are collected in Table 5.

From 1999 onwards, one finds that most of the analysis has been done by mathematicians focusing on the nature of space [55-61], self-adjointness of operators [62-65], Hardy-Dirac inequalities [66-71], and domains of validity of the min-max theorem for systems with different potentials [72]. These issues are generally of less interest in science as they lack direct physical significance. Reference 61 is a review, and it can introduce any willing physical scientist into the world of mathematical art with heightened perspectives.

### 4.5. Missing Items

The reader needs to realize that the references given here are by no means complete, and that almost all the references quoted (barring a few earlier ones) relate to the variation problem. For example, no discussion has been done on methods that are derived from an application of the Foldy-Wouthuysen transformation and similar transformations and rely on the two-component trial spinor. Methods that use perturbation theory and many-body methodologies are more or less ignored unless a contribution is historically important. A description of RCI calculations is avoided here, as it is of no interest in the present case unless it can shine new light on how the RCI basis can be restricted to the positive-energy sector. Issues like parity non-conservation in atoms, and the "constraint" formalism to achieve a covariant equation of motion are not relevant here. In short, almost all literature on relativistic atomic and molecular calculations done after the establishment of the MMP is absent here. For a wider data base on the general subject, one may consult the compiled volume by Pyykko [73]. A general understanding of methodologies can be obtained from the review articles by Mohanty and his coworkers [74-76]. Incidentally, the book edited by Clementi [74-76] is a wonderful source of knowledge on the development of computational chemistry methodologies and programming such as ATOMCI, ALCHEMY II, KGNMOL, MELD, MOLCAS, AMPAC, HONDO, HCOIN, SIRIUS, RMPROP, BNDPKG2, interactive visualization techniques, and LCAP. General theoretical knowledge can be gleaned from the book by Lindgren and Morrison [77]. Chapter 3 and Chapter 4 of this research monograph offer a good exposition of angular momentum graphs. This book is mainly on the non-relativistic theory, though it discusses the general theory and methodology in detail. Hyperfine interactions and relativistic effects are discussed only in Chapter 14.

### 5. Discussion

### 5.1. Limitations – Negative energy solutions

The min-max theorem is unambiguously restricted to cases where an energy gap exists between the discreet eigenvalues and the eigenvalues of negative mass. It may not hold in the zero mass limit. It is valid for the first solution for energy in (18) for which $+mc^2 > \varepsilon_+^0[u] > 0$ when $\alpha Z < 1$, and does not apply to the second solution. It has been already shown in equations (28) through (30) that a max-min principle is operative for $\varepsilon_-^0[u]$.

### 5.2. Other limitations

Limitations in the form of the self-adjointness of Dirac operators, Hardy-like inequalities and the variable domain of Coulomb potential have been discussed [62-72]. These discussions are mostly of interest to mathematicians and not concerned with the physical world, and often known but not stated in mathematical language. For instance, in ref. [72] Esteban, Lewin, and Séré considered Dirac operators with a Coulomb-type potential V (x) ~ −ν/|x| (with $mc^2$ as the unit of energy) that has a strong Coulomb singularity ($3^{1/2}/2 \leq \nu \leq 1$). It was shown that the operator has a distinguished self-adjoint extension. Furthermore, the min-max formulas are shown to be valid for the eigenvalues in the gap, in simple function spaces that are independent of the value of ν in the range $0 \leq \nu \leq 1$.

### 5.3. Mathematicians' Dirac-Coulomb

Another area of confusion exists. When a physical scientist refers to Dirac-Coulomb, he or she normally attributes it to the marriage of Dirac's Hamiltonian for each individual spin-1/2 fermion and the bare Coulomb interaction between each pair of indistinguishable particles. So conventionally Dirac-Coulomb equation is referred to systems with two or more electrons with $e^2/r_{12}$ terms. The interaction is at zero order QED, neither made covariant by adding Gaunt term on an ad-hoc basis nor found retarded as deduced by Breit [2], and it is certainly not projected as suggested by Sucher [3-4]. A quantum chemical calculation using normalizable basis functions makes it in principle projected as pointed out by Sucher. The min-max treatment makes it effectively projected.

In mathematical parlance, the term Dirac-Coulomb has been applied mostly for hydrogen-like systems. The mathematician's logic is that one can always describe the movement of a spin-1/2 fermion in any arbitrary external potential $V_{ext}$ such as the Kronig-Penny potential in solid state theory. Hence in mathematics, Dirac-Coulomb means Dirac's one-electron equation with Coulomb potential as the external potential $V_{ext}$. Also, the external potential may be scaled, thereby giving rise to exotic treatments like Dirac-scaled-Coulomb, etc., areas that are prime candidates for an investigation of the validity of the min-max theorem [31, 65, 70-72].

### 5.4. Solution of Klein-Gordon equation

The present discussion remains incomplete without a brief consideration of the solution of the relativistic wave equation for a spin-0 boson. The latter equation, suggested by Klein and Gordon, is a follow-your-nose quantum mechanical version $-\pi_\mu \pi^\mu \Psi = m^2 c^2 \Psi$ of the classical relativistic kinematics for one particle moving in the presence of 4-potential. It is generally written in the form of second order in time.

Hall and his co-workers have extensively investigated the time-independent problem. For example, Hall studied the non-local interactions and the *N*-boson problem [78], and the semi-relativistic *N*-boson systems bound by attractive pair potentials [79]. Hall et al. analysed relativistic *N*-boson systems bound by pair potentials [80-81] and went on to establish two comparison theorems for a central potential. Hall and Lucha considered the semi-relativistic stability of similar systems [82]. Hall and Aliyu studied the comparison theorems for the Klein-Gordon equation in *d* dimensions [83]. Hall also considered comparison theorems for solving Dirac's equation [84], though his interest mainly involved aspects related to supersymmetry.

The single-particle Klein–Gordon equation can also be written as first-order in time, with the wave function described by a two-component column matrix. It is well-known that the two components are related to each other. The optimum upper component–lower component coupling operator was found for a trial function representing a bound state for an attractive potential [85]. It corresponds to an energy minimum, instead of a maximum as in the case of Dirac operator. A further variation of the upper component leads to a min–min theorem [85]. Besides, the two comparison theorems put forward by Hall et al. for the solution of the second-order Klein–Gordon equation for a particle moving in an attractive central potential [81-83] could be verified from the two-component approach. A relation was also derived in the case a uniform magnetic field is switched on. An explicit discussion was given for a Coulomb potential [85].

**Acknowledgments**

I am grateful to W. T. Borden who had encouraged me some time back to write a review in relativistic quantum chemistry, and to A. K. Pal who has helped with the preparation of the figures.

# TABLES

**Table 1.** Application of min-max theory and principle, and alternative techniques: solution of Dirac equation for hydrogen-like systems and one-electron molecule-ions

| Example | Subject | Reference | Year |
|---|---|---|---|
| | Dirac equation for hydrogen-like systems | | |
| Discrete basis | Analytical results were derived, and spurious roots were noted. | 12 | 1981 |
| Calculations: choice of basis sets | A quasi-unitary transformation of the Dirac equation in spinor space results in better behaved matrix representations. | 13-14 | 1982 |
| Discrete basis | Results of [12] were reproduced for same radial components of $u$ and $l$. | 19 | 1984 |
| Discrete basis | Relativistic virial relation | 39 | 1984 |
| Hydrogenic atoms | Choice of basis set, and minimal basis calculations | 23 | 1984 |
| Basis expansion | Basis set derived from STOs | 26 | 1986 |
| Discrete basis | Relativistic extension of Hohenberg-Kohn theorem | 40 | 1987 |
| Finite basis set | H atom solutions from Sturmian-type basis set | 28 | 1989 |
| Discrete basis | Analytical results | 41 | 1992 |
| Finite basis set | General discussion | 33 | 1997 |
| Discrete basis | General discussion | 31 | 1999 |
| Discrete basis | General discussion on application to Dirac operator with a Coulomb-like potential: the result is optimal for the Coulomb potential. | 32 | 2000 |
| Different bases | General operator technique | 38 | 2000 |
| | Dirac equation for one-electron molecule-ions | | |
| $H_2^+$ | Minimal basis derived from STOs | 19 | 1984 |
| $H_2^+$ | Basis set derived from STOs: collapse illustrated by "Shower" diagrams | 27 | 1988 |
| $H_2^+$ | Basis set derived from STOs | 42 | 1992 |

**Table 2.** Application of min-max theory and principle and alternative techniques: solution of Dirac-Fock equations for atoms and molecules

| Example | Subject | Reference | Year |
|---|---|---|---|
| | Dirac-Fock equations for atoms and molecules | | |
| Be, $Be_2$ | With negative-energy correction, basis set derived from GTOs | 43 | 1982 |
| Be | Discrete basis set | 44 | 1983 |
| General atoms | Matrix representation of operator products | 22 | 1984 |
| $Z = 1-10$ | Kinetically balanced basis set derived from GTOs | 34 | 1985 |
| General atoms | Partitioning technique | 24 | 1985 |
| General atoms | Atomic calculations | 25 | 1986 |
| $H_2$, He | Basis set derived from STOs | 27 | 1988 |
| He | Basis set derived from STOs | 29 | 1992 |
| H, Li, Be, LiH, $Be_2$ | Basis sets derived from GTOs | 35 | 1993 |
| $Z = 1-10$ | Optimized coupling basis set derived from Partridge Gaussian functions | 36 | 1995 |
| He-like ions | Basis set derived from STOs | 30 | 1996 |

**Table 3.** Application of min-max theory and principle and alternative techniques: General Dirac-Fock and many-body calculations and Dirac-Coulomb calculations on two-electron atoms

| Example | Subject | Reference | Year |
|---|---|---|---|
| General Dirac-Fock and many-body calculations using MMP or related principles | | | |
| Be atom, $Be_2$ molecule | General molecular treatment implementing negative energy correction by orthogonality constraint | 43 | 1982 |
| General molecule | Kinetic balance and basis set expansion | 13-14 | 1982 |
| H | General methodology for atomic calculations using the proper matrix representation for kinetic energy | 22-23 | 1984 |
| General atoms | General methodology for partitioning technique: widely practised today | 24 | 1985 |
| Atoms with $Z \leq 10$ | Relativistic Gaussian basis sets: large and small component fitting | 34 | 1985 |
| H, Li, Be, LiH, $Be_2$ | Basis sets derived from GTOs | 35* | 1993 |
| Atoms with $Z \leq 10$ | Relativistic Gaussian basis sets: upper component built from Patridge basis set; lower component prepared by optimizing the coupling operator of form $\lambda \sigma \cdot \mathbf{p}$ | 36 | 1995 |
| Dirac-Coulomb calculations on two-electron atoms | | | |
| He-like systems | DF, (Positive-energy) RCI, Min-max (All-energy RCI) and QED calculations using basis sets prepared from STOs with varying exponents | 29 | 1992 |
| He-like systems | Varying nonlinear parameters and basis sets prepared from STOs | 30 | 1996 |

**Table 4.** Earlier molecular calculations without any variation safeguard

| Topic | System | Reference | Year |
|---|---|---|---|
| Very simple LCAS-MS scheme | $H_2^+$ | 45 | 1967 |
| Perturbative calculation on $1s\sigma_g$ state | $H_2^+$ | 46 | 1969 |
| Symmetry of diatomic molecular spinors | Diatomics | 47 | 1974 |
| The earliest proposal of closed-shell molecular treatment | Diatomic molecular integrals | 48 | 1975 |
| Symmetry of polyatomic molecular spinors | Polyatomics | 49 | 1976 |
| Symmetry of polyatomic molecular spinors | Polyatomics | 50 | 1976 |
| The earliest proposal of open-shell molecular treatment | Molecular integrals | 51 | 1980 |
| One-centre Dirac-Fock numerical calculation | BH, AlH, GaH, InH, TlH | 52 | 1976 |
| LCAS-MS calculation | AgH, AuH | 53 | 1982 |

**Table 5.** Miscellaneous application of min-max theory and principle and alternative techniques

| Application | System | Reference | Year |
|---|---|---|---|
| Relativistic virial theorem in terms of the scale-optimized upper component function in non-relativistic limit | Electron in a central field | 18 | 1975 |
| Negative mass correction | Pb atom | 8 | 1980 |
| Effective Hamiltonian | General | 11 | 1981 |
| Negative mass correction | Be atom, $Be_2$ | 43 | 1982 |
| Relativistic virial theorem in terms of 4-component spinors | Hydrogen-like system | 39 | 1984 |
| Variation analysis | Review | 25 | 1986 |
| Relativistic extension of Hohenberg-Kohn theorem | Hydrogen-like system | 40 | 1987 |
| Solutions for different angular momentum, variation collapse and continuum state | Relevant examples | 41 | 1992 |
| Min-max theorem and principle | Review | 54 | 1994 |
| A critical assessment | Operator technique | 38 | 2000 |

**References**
[1] G. E. Brown, D. G. Ravenhall, *Proc. Royal Soc. London* **1951**, *A208*, 552-559.
[2] G. Breit, *Phys. Rev.* **1929**, *34*, 553-573.
[3] J. Sucher, *Phys. Rev.* **1980**, *A22*, 348-362; Erratum: **1981**, *A23*, 388.
[4] J. Sucher, *Phys. Scripta* **1987**, *36*, 271-281.
[5] I. P. Grant, *Adv. Phys.* **1970**, *19*, 747-811.
[6] J.-P. Desclaux, *Comput. Phys. Commun.* **1975**, *9*, 31-45.
[7] D. R. Hartree, The Calculation of Atomic Structures; Wiley: New York, **1957**.
[8] S. N. Datta, *Chem. Phys. Letters* **1980**, *74*, 568-572.
[9] J. D. Weeks, A. Hazi, S. A. Rice, *Adv. Chem. Phys.* **1969**, *16*, 283-342.
[10] H. Wallmeier, W. Kutzelnigg, *Chem. Phys. Letters* **1981**, *78*, 341-346.
[11] Y. Ishikawa, G. L. Malli, *Chem. Phys. Letters* **1981**, *80*, 111-118.
[12] G. W. F. Drake, S. P. Goldman, **Phys. Rev.** **1981**, *A23*, 2093-2098.
[13] F. Mark, W. H. E. Schwarz, *Phys. Rev. Letters* **1982**, *48*, 673-676.
[14] W. H. E. Schwarz, H. Wallmeier, *Mol. Phys.* **1982**, *46*, 1045-1061.
[15] W. H. E. Schwarz, E. Wechsel-Trakowski, *Chem. Phys. Letters* **1982**, *85*, 94-97.
[16] B. Gazdy, *Chem. Phys. Letters* **1983**, *99*, 41-44.
[17] B. Gazdy, K. Ladanyi, *J. Chem. Phys.* **1984**, *80*, 4333-4340.
[18] F. Rosicky, F. Mark, *J. Phys. B – At. Mol. Phys.* **1975**, *8*, 2581-2588.
[19] S. N. Datta, S. Jagannathan, *Pramana – J. Phys.* **1984**, *23*, 467-473.
[20] R. E. Stanton, S. Havriliak, *J. Chem. Phys.* **1984**, *81*, 1910-1918.
[21] K. G. Dyall, *Chem. Phys. Letters* **1990,** *174*, 25-32.
[22] K. G. Dyall, I. P. Grant, S. Wilson, *J. Phys. B – At. Mol. Phys.* **1984**, *17*, L45.
[23] K. G. Dyall, I. P. Grant, S. Wilson, *J. Phys. B – At. Mol. Opt. Phys.* **1984**, *17*, 493-504.
[24] J. Wood, I. P. Grant and S. Wilson, *J. Phys.* **1985**, *B18*, 3027-3042.
[25] I. P. Grant, *J. Phys.* **1986**, *B19*, 3187-3206.
[26] J. D. Talman, *Phys. Rev. Lett.* **1986**, *57***,** 1091-1094.
[27] S. N. Datta, G. Devaiah, *Pramana – J. Phys.* **1988**, *30*, 387-405. Also see: Abstract of the Adriatico Research Conference on Relativistic Many-Body Effects, Trieste, **1986**.
[28] G. W. F. Drake, S. P. Goldman, *Adv. Atomic Molecular Phys.* **1989**, *25*, 393-416.
[29] S. N. Datta, *Pramana – J. Phys.* **1992**, 38, 51-75.
[30] A. Kolakowska, J. D. Talman, K. A. Ashamar, *Phys. Rev.* **1996**, *A53*, R168-R177.

[31] M. Griesemer, R. T. Lewis, H. Siedentop, *Doc. Math.* **1999**, *4*, 275-283.
[32] J. Dolbeault, M. J. Esteban, É. Séré, *J. Funct. Anal.* **2000**, *174*, 208-226.
[33] W. Kutzelnigg, *Chem. Phys.* **1997**, *225*, 203-222.
[34] F. Mark, Schriftenreihe des Max-Planck-Instituts für Strahlenchemie, No. 24; MPI: Mülheim, **1985**.
[35] S. N. Datta, *Pramana – J. Phys.* **1993**, *41*, 363-370.
[36] S. N. Datta, *Int. J. Quantum Chem.* **1995**, *56*, 91-95.
[37] H. Partridge, NASA Technical Memorandum, No. 101044; NASA: Moffett Field, **1989**.
[38] S. N. Datta, *Pramana – J. Phys.* **2000**, *55*, 383-392.
[39] S. N. Datta, *Pramana – J. Phys.* **1984**, *23*, L275-L278.
[40] S. N. Datta, *Pramana – J. Phys.* **1987**, *28*, 633-639.
[41] S. Datta, S. N. Datta, *Pramana – J. Phys.* **1992**, *38*, 521-530.
[42] L. LaJohn, J.D. Talman, *Chem. Phys. Letters* **1992**, *189*, 383-389.
[43] S. N. Datta, C. S. Ewig, *Chem. Phys. Letters* **1982**, *85*, 443-446.
[44] Y. Ishikawa, R. C. Binning Jr., K. M. Sando, Chem. Phys. Letters **1983**, *101*, 111-114.
[45] P. I. Pavlik, S. M. Blinder, *J. Chem. Phys.* **1967**, *46*, 2749-2751.
[46] S. K. Luke, G. Hunter, R. P. McEachran, M. Cohen, *J. Chem. Phys.* **1969**, *50*, 1644-1650.
[47] J. Oreg, G. Malli, *J. Chem. Phys.* **1974**, *61*, 4349-4356.
[48] G. Malli, J. Oreg, *J. Chem. Phys.* **1975**, *63*, 830-841.
[49] J. Oreg, G. Malli, *J. Chem. Phys.* **1976**, *65*, 1746-1754.
[50] J. Oreg, G. Malli, *J. Chem. Phys.* **1976**, *65*, 1755-1763.
[51] G. Malli, *Chem. Phys. Letters* **1980**, *73*, 510-513.
[52] P. Pyykkö, J. P. Desclaux, *Chem. Phys. Letters* **1976**, *42*, 545-549.
[53] Y. S. Lee, A. D. McLean, *J. Chem. Phys.* **1982**, *76*, 735-736.
[54] S. N. Datta, *Proc. Ind. Acad. Sci. – Chemical Sciences* **1994**, *106*, 445-466.
[55] M. Griesemer, H. Siedentop, *J. London Math. Soc.* **1999**, *60*, 490–500.
[56] J. Dolbeault, M. J. Esteban, É. Séré, *J. Funct. Anal.* **2000**, *174*, 208–226.
[57] J. Dolbeault, M. J. Esteban, É. Séré, *Calc. Var. Partial Differ. Equ.* **2000**, *10*, 321–347.
[58] J. Dolbeault, M. J. Esteban, É. Séré, M. Vanbreugel, *Phys. Rev. Lett.* **2000**, *85*, 4020–4023.
[59] J. Dolbeault, M. J. Esteban, É. Séré, *Int. J. Quantum Chem.* **2003**, *93*, 149 –155.
[60] J. Dolbeault, M. J. Esteban, É. Séré, *J. Eur. Math. Soc.* **2006**, *8*, 243–251.
[61] M. J. Esteban, M. Lewin, É. Séré, *Bull. Amer. Math. Soc.* (*N.S.*) **2008**, *45*, 535–593.
[62] W. D. Evans, *Proc. London Math. Soc.* **1970**, *20*, 537–557.
[63] J. J. Landgren, P. A. Rejto, *J. Math. Phys.* **1979**, *20*, 2204–2211.
[64] J. J. Landgren, P. A. Rejto, M. Klaus, *J. Math. Phys.* **1980**, *21*, 1210–1217.
[65] H. Hogreve, *J. Phys. A – Math. Gen.* **2013**, *46*, 025301.
[66] J. Dolbeault, M. J. Esteban, M. Loss, L. Vega, *J. Funct. Anal.* **2004**, *216*, 1-21.
[67] J. Dolbeault, M. J. Esteban, J. Duoandikoetxea, L. Vega, *Ann. Sci. Ecole Norm. Sup.* **2007**, *40*, 885 – 900.
[68] M. J. Esteban, M. Loss, *J. Math. Phys.* **2007**, *48*, 112107-112108.
[69] M. J. Esteban, M. Loss, in Mathematical Results in Quantum Mechanics; I. Beltita, G. Nenciu, R. Purice, Eds.; World Sci.: Hackensack, NJ, **2008**; pp. 41–47.
[70] S. Morozov, D. Müller, *Math. Z.* **2015**, *280*, 733–747.
[71] D. Müller, Minimax principles, Hardy-Dirac inequalities and operator cores for two and three dimensional Coulomb-Dirac operators, arXiv:1603.01557v1 [math-ph], **2016**.
[72] M. J. Esteban, M. Lewin, É. Séré, Domains for Dirac-Coulomb Min-Max levels, https://arxiv.org/pdf/1702.04976, **2017**.
[73] P. Pyykkö, Relativistic Theory of Atoms and Molecules: A Bibliography 1916-1985; Lecture Notes in Chemistry, Vol. 41; Springer: Berlin, **1986**.

[74] A. K. Mohanty, F. A. Parpia, E. Clementi, in Modern Techniques in Computational Chemistry: MOTECC-91; E. Clementi, Ed.; ESCOM: Leiden, 1991; Chapter 4.
[75] F. A. Parpia, A. K. Mohanty, in Modern Techniques in Computational Chemistry: MOTECC-91; E. Clementi, Ed.; ESCOM: Leiden, 1991; Chapter 5.
[76] A. K. Mohanty, S. Panigrahy, E. Clementi, in Modern Techniques in Computational Chemistry: MOTECC-91; E. Clementi, Ed.; ESCOM: Leiden, 1991; Chapter 15.
[77] I. Lindgren, J. Morrison, Atomic many-body theory, 2nd Edition; Springer: Berlin, **1985**.
[78] R. L. Hall, *Z. Physik* **1979,** *A291*, 255-259.
[79] R. L. Hall, *J. Phys.* **1989**, *A22*, 179-183.
[80] R. L. Hall, W. Lucha, F. F. Schöberl, *J. Math. Phys.* **2002**, *43*, 1237-1246; Erratum, **2003,** *44*, 2724.
[81] R. L. Hall, W. Lucha, F. F. Schöberl, *J. Math. Phys.* **2004**, *45*, 3086-3094.
[82] R. L. Hall, W. Lucha, *J. Phys. A* **2008**, *41*, 355202.
[83] R. L. Hall, M. D. Aliyu, *Phys. Rev.* **2008**, *A78*, 052115.
[84] R. L. Hall, *Phys. Rev.* **2010**, *A81*, 052101.
[85] S. N. Datta, A. Ghosh, R. Chakraborty, *Ind. J. Phys.* **2015**, *89*, 181-187.